*Article*

# Cross-Layer Adaptive Feedback Scheduling of Wireless Control Systems

Feng Xia [1,4], Longhua Ma [2,*], Chen Peng [3], Youxian Sun [2] and Jinxiang Dong [1]

1  College of Computer Science and Technology, Zhejiang University, Hangzhou 310027, China
2  State Key Lab of Industrial Control Technology, Zhejiang University, Hangzhou 310027, China
3  School of Electrical and Automation Engineering, Nanjing Normal University, Nanjing 210042, China
4  Faculty of Information Technology, Queensland University of Technology, Brisbane QLD 4001, Australia

* Author to whom correspondence should be addressed; E-Mails: f.xia@ieee.org (F.X.); lhma@iipc.zju.edu.cn (L.M.)

**Abstract:** There is a trend towards using wireless technologies in networked control systems. However, the adverse properties of the radio channels make it difficult to design and implement control systems in wireless environments. To attack the uncertainty in available communication resources in wireless control systems closed over WLAN, a cross-layer adaptive feedback scheduling (CLAFS) scheme is developed, which takes advantage of the co-design of control and wireless communications. By exploiting cross-layer design, CLAFS adjusts the sampling periods of control systems at the application layer based on information about deadline miss ratio and transmission rate from the physical layer. Within the framework of feedback scheduling, the control performance is maximized through controlling the deadline miss ratio. Key design parameters of the feedback scheduler are adapted to dynamic changes in the channel condition. An event-driven invocation mechanism for the feedback scheduler is also developed. Simulation results show that the proposed approach is efficient in dealing with channel capacity variations and noise interference, thus providing an enabling technology for control over WLAN.

# 1. Introduction

With recent advances in wireless technologies, wireless control systems (WCSs) are attracting increasing attention from both academia and industry [1-4]. In a WCS, spatially distributed nodes of sensors, controllers and/or actuators are interconnected with wireless links. The use of wireless technologies in control applications has many advantages compared to wired networked control systems that are dominant at the moment. For instance, wireless networks allow flexible installation and maintenance, mobile operation, and monitoring and control of equipments in hazardous and difficult-to-access environments. Another important factor that instigates the deployment of wireless sensor/actuator networks is their relatively cheaper costs.

However, wireless communications raise new challenges for control system analysis and design. Wireless channels have adverse properties, such as path loss, multi-path fading, adjacent channel interference, Doppler shifts, and half-duplex operations [1]. While traditional wired networks usually have fixed communication capacity, the link capacity of wireless channels may vary significantly over time [5-7]. Because the operations of wireless transceivers are half-duplex, wireless systems cannot support non-destructive medium access control (MAC) protocols. From the control point of view, communication networks introduce problems related to delay, packet losses, and jitters. Compared with wirelines, wireless links make these problems more pronounced [8,9]. For instance, the bit error rate of a wireless channel is typically several times higher than that of a wired connection [10]. These phenomena degrade the quality of control (QoC), or even cause system instability in extreme circumstances [5,11].

The area of WCSs is still in its infancy. The suitability of diverse wireless technologies for control applications has been studied through both simulations [12-14] and experiments [7,10,15]. A number of proposals on modifying established communication mechanisms for wireless networks to achieve real-time guarantees have been presented, e.g. [16,17]. Some other researchers, mostly from the control community, attempt to design controllers robust to the temporal non-determinism of wireless networks, for example, [6,18].

In contrast to all these papers, the focus of this work is on co-design of real-time control and wireless communications. Because of its interdisciplinary nature, this co-design is complicated, with limited results reported in the literature. Liu and Goldsmith [19] introduced the methodology of cross-layer design into WCS design, and presented a four-layer framework. But adaptation of the sampling periods of control loops is not considered. Through studying the impact of varying fading wireless channels on control performance, Mostofi and Murray [5] suggested that the controller parameters should be dynamically adapted with respect to channel conditions. An offline approach to optimize the stationary performance of a linear control system by jointly allocating communication resources and tuning parameters of the controller is presented in [20]. Different methods for adapting sampling periods at runtime have been developed in e.g. [10,11,21]. All these methods are based on algorithms with fixed parameters. Consequently, the effects of varying channel conditions such as changes in network transmission rates are not attacked. In our recent work [3,4,9], we presented several design methods for control systems over wireless networks. An integrated framework that adjusts the maximum number of allowable data retransmission attempts and tunes the controller parameters is given in [22]. Different approaches to dynamic bandwidth allocation through dynamically adjusting sampling periods are presented in [23,24] for wireline networked control systems. Additionally, almost all existing solutions for online sampling period adjustment are time triggered.

Considering WCSs closed over IEEE 802.11b WLAN, this paper develops a cross-layer adaptive feedback scheduling (CLAFS) scheme [25] that dynamically adjusts the sampling periods with respect to variable channel conditions. The primary objective is to provide QoC guarantees for WCSs via flexible resource management in dynamic environments that feature noise interference and variability of the network transmission rate. Based on cross-layer design, this scheme takes advantage of sharing and exchanging of information across the physical layer and the application layer within the communication protocol stack of WLAN. The sampling periods of control loops are adapted online to control the deadline miss ratio (DMR). To cope with dynamic variations of the link capacity, the feedback scheduler uses a simple proportional control algorithm with adaptive parameters. Since interference and node movement in wireless systems are stochastic and unpredictable in most situations, it is usually hard to select an appropriate invocation interval for a time-triggered feedback scheduler. To address this problem, an event-driven invocation mechanism for the feedback scheduler is suggested. This mechanism contributes not only to reduction of overheads (on average), but also to quick responses to changes in communication resource availability, resulting in further improvement of practical performance of the feedback scheduler.

This paper is organized as follows. Section 2 describes the architecture of the control system considered. The case used as an illustrative example throughout the paper is also given. In Section 3, the employed cross-layer design methodology is described, followed by an analysis of the temporal properties of the studied WCS in terms of DMR. Then the CLAFS scheme is presented along with relevant algorithms. Section 4 presents the event-driven invocation mechanism for the feedback scheduler. In Section 5 the effectiveness of the proposed approach is validated by simulations highlighting its advantages. Finally, Section 6 concludes with discussions on possible extensions over the proposed approach.

## 2. System Model

Consider a WCS shown in Fig. 1, where, besides an interfering loop, there are altogether $N$ independent control loops. Each control loop consists of a smart sensor (S), a smart actuator (A), a controller (C) and a physical process (P) to be controlled. To facilitate time synchronization, assume that the sensor and the actuator run on top of the same clock platform. The nodes communicate using the IEEE 802.11b protocol. The computation times of all control tasks on the controllers are assumed to be negligible relative to communication delays. The total delay within a control loop is consequently equal to the sum of the communication delay of sampled data from the sensor to the controller and the communication delay of control command from the controller to the actuator, including both waiting delays and transmission delays.

In the context of wireless control, there are basically two classes of deadline misses. The first class is that the sample data or the control command is truly lost during the course of transmission due to bit errors, noise interference, low received signal strengths, etc. In contrast, in the second class of deadline misses, the control command is actually received by the actuator, but the communication delay exceeds the deadline, which equals the sampling period.

**Figure 1.** Architecture of a wireless control system.

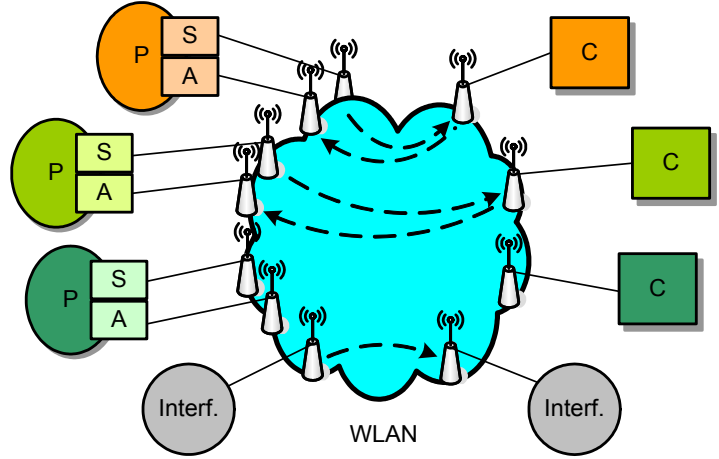

*2.1. Communication over WLAN*

IEEE 802.11b protocol [26] specifies two medium access coordination functions, the mandatory distributed coordination function (DCF) that is based on Carrier Sense Multiple Access with Collision Avoidance (CSMA/CA) and the optional point coordination function (PCF). Unlike wired nodes, wireless nodes cannot detect collisions because they are half-duplex, i.e. they cannot send and receive signals at the same time. CSMA/CA delivers a best effort service, thereby providing no bandwidth and delay guarantees.

In IEEE 802.11, each node senses the medium before starting a transmission. If the medium is idle for at least a DCF interframe space (DIFS), the packet is transmitted immediately. If the medium is sensed busy, the node waits for the end of the current transmission and then starts the contention, also called backoff process. The node selects a random backoff time. During the backoff process, the backoff timer is decremented in terms of slot time as long as the medium is idle. When the medium is busy, the timer is frozen. When its backoff timer expires, if the network is still idle, the data packet is sent out. The node having the shortest contention delay wins and transmits its packet. The other nodes just wait for the next contention. If another collision occurs, a new backoff time is chosen and the backoff procedure starts over again until some time limit is exceeded.

*2.2. Case Study*

There are three identical control loops in the WCS, i.e. $N = 3$. Each of the processes under control is an independent DC motor [27] modelled in continuous-time form:

$$G(s) = \frac{2029.826}{(s+26.29)(s+2.296)} \quad (1)$$

The controllers use the PID (Proportional-Integral-Derivative) control law with a continuous-time form $G_{PID}(s) = K_P + \frac{K_I}{s} + K_D s$. The controller parameters are: $K_P = 0.1701$, $K_I = 0.378$, and $K_D = 0$. Digital controllers are designed by discretizing continuous-time controllers. As sampling periods are changed, the controller parameters of digital controllers are updated accordingly.

Due to node movement, the communication distance between the controller and the process (where the sensor and the actuator are attached) may change during runtime. According to the properties of wireless signal transmission, the received signal strengths drop with increasing communication distances. When the signal-to-noise ratio of the received signals is below a certain level, IEEE 802.11b will make the trade-off between speed and communication reliability by reducing the transmission rate, for example, from the usual maximum value of 11 to 5.5, 2, or even 1 Mb/s [10]. This inherent feature of 802.11b gives rise to variability of channel capacity, a crucial issue that should be taken into account when designing control systems closed over WLAN.

Apart from the changes in channel capacity, another problem that needs to be addressed is the effect of noise interference on QoC. In the subsequent sections, a general solution for these problems will be proposed and validated, while using this case as an illustrative example.

## 3. Cross-Layer Adaptive Feedback Scheduling

To enable wireless control in dynamic environments, the feedback scheduling technology is adopted. It has been shown that feedback scheduling is quite promising in managing uncertainties in resource availability [28,29]. This motivates the use of this technology in dynamic management of the variable communication resources in WLAN. To cope with the adverse properties of wireless communications, the cross-layer design methodology, a technique that is gaining increasing importance in networking applications, is incorporated with feedback scheduling.

*3.1. Cross-Layer Design Methodology*

The design of wireless networks is often based on a layered network protocol stack, and the design and operation of different network layers are separated. As shown in Fig. 2, IEEE 802.11b specifies two layers, i.e., the physical layer and the MAC sub-layer, among the seven-layer OSI reference model. At the physical layer 802.11b specifies four different levels of transmission rates, i.e., 1, 2, 5.5, and 11Mb/s. At the MAC layer 802.11b exploits CSMA/CA to solve resource contention among multiple nodes. In the context of wireless control, it is intuitive that the control applications are at the application layer.

**Figure 2.** Cross-layer design framework for wireless control systems.

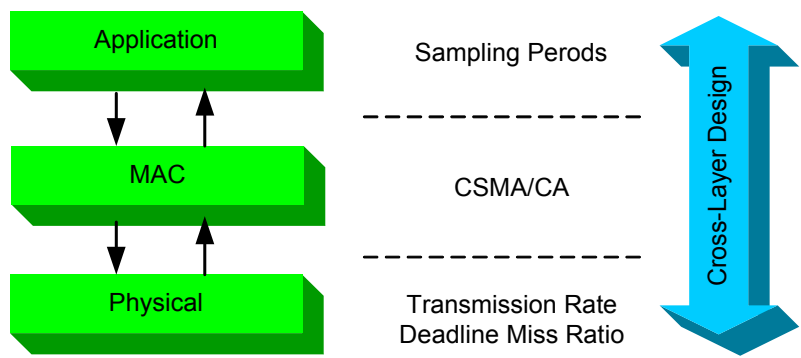

When the system operates under dynamic environments, the timing properties of WLAN may vary with different physical layer parameters, which influence in turn the performance of the control systems at the application layer. To maximize QoC, it is necessary to take advantage of dynamic interactions between the physical layer and the application layer. Cross-layer design should be exploited to achieve application adaptation [19,30].

In this paper, the sampling periods of control loops are chosen as the parameters of the application layer. The main reason behind this choice is that the sampling periods influence not only the QoC but also the workload on the network, which affects the communication delay and the DMR. In a sense, the DMR can be regarded as an indicator for link quality associated with the physical layer. Since the transmission rate may change at runtime, it naturally becomes another parameter at the physical layer. Consequently, the basic role of feedback scheduling that exploits cross-layer design is to adjust the sampling periods of control systems at the application layer based on information about DMR and transmission rate from the physical layer.

In wireless networked systems, a straightforward design objective of feedback scheduling is to control the DMR at a desired level. Since WLAN does not support non-destructive communication protocols, it is impossible to analyse the system schedulability for WCSs. Therefore, the network utilization is not a suitable choice for the controlled variable for feedback scheduling. Without loss of generality, the DMR of all control loops is used as the controlled variable of the feedback scheduling system. Actually, because WLAN employs a MAC protocol featuring random medium access, the DMR of control loops also reflects the level of DMR of interfering signals.

*3.2. Analysis of Deadline Misses over WLAN*

Before designing the feedback scheduling algorithm, the temporal behaviour of WLAN in terms of DMR needs to be studied. In the following, the effects of the transmission rate $r$ and the sampling period $h$ on the DMR $\rho$ are analysed through simulation experiments.

**Figure 3.** Deadline miss ratio of the wireless control system.

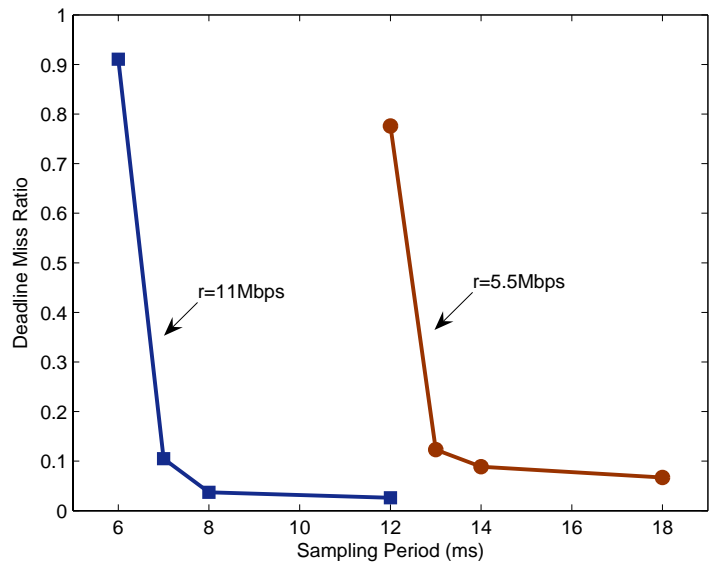

Assume that there is no interfering signal in the system shown in Fig. 1, and the sizes of all data packets to be transmitted over the network are 1 KB. Fig. 3 depicts the DMR of the system with different transmission rates and different sampling periods. For each pair of transmission rate and sampling period, the simulation is run 10 times, each for 3 seconds. Each value given in Fig. 3 is the arithmetical mean of the DMRs recorded separately in these 10 runs.

Since the three control loops in the system are completely identical, every change in sampling period shown in Fig. 3 implies that the sampling periods of all three control loops are adjusted at the same time. With a given transmission rate, the DMR decreases as the sampling period increases, and the rate of the change also decrease gradually.

When the sampling period is relatively small, the network traffic is heavy, resulting in frequent collisions among communication nodes. As a result, the communication delay of a data packet may become large, even go beyond the deadline, or the data packet may be dropped due to too many retransmission attempts. In such situations, the DMR will be large. Enlarging the sampling period can reduce the DMR. The reasons behind can be explained as follows.

- Firstly, the increase of sampling period reduces the amount of network traffic and hence the probability of node collisions. Consequently, the communication delay of data packets decreases on average, and the possibility of data packets being dropped also decreases thanks to the reduction in the number of retransmission attempts.
- Secondly, as the sampling period increases, the deadlines of data packets to be transmitted increase accordingly. Consequently, longer communication delays are allowed. Both of these effects result in reduced DMRs.

Comparing the results for $r = 5.5$ Mbps and $r = 11$ Mbps, it can be seen that larger transmission rate benefits the reduction of DMR, especially when the communication resources are scarce. For instance, when h = 12 ms, the DMRs for $r = 5.5$ and 11 Mbps are $\rho = 77.6\%$ and 2.7%.

## 3.3. Adaptive Feedback Scheduling Algorithm

From the above analysis, the basic idea of feedback scheduling of WCSs can be stated as follows: with the goal of maximizing QoC, dynamically adjust the sampling periods of control loops to maintain the DMR at a desired level. From the control perspective, lower DMRs are always better. Therefore, large sampling periods should be used to avoid deadline misses.

However, it is not easy to completely avoid deadline misses in a typical wireless environment. As also shown in Fig. 3, in order to reduce the DMR to a near-zero level, quite a large sampling period has to be assigned to each control loop. Unfortunately, according to sampled-data control theory, such a large increase in sampling period could degrade the QoC remarkably. In this context, the resulting QoC of the system may be adversely deteriorated, regardless of the decrease in the DMR. Therefore, in WCSs it is favourable to maintain the DMR at an appropriate non-zero level [10,21].

Within the framework of feedback scheduling, we use a simple proportional control algorithm to adjust the sampling period:

$$\Delta h(j) = K \cdot e(j) \tag{2}$$

where $K$ is the proportional coefficient, $e(j)=\rho(j)-\rho_r$ is the difference between actual DMR and its desired value, and $j$ is the index of the invocation instant of the feedback scheduler. Taking into account the constraint on the maximum allowable sampling period $h_{max}$ of the control loops, the sampling period at $j$-th instant is then calculated by:

$$h(j) = \min\{h(j-1) + K \cdot e(j), h_{\max}\} \tag{3}$$

In the above algorithm, the proportional coefficient $K$ and the DMR setpoint $\rho_r$ are two key design parameters. Related design considerations are described below.

3.3.1. Proportional Coefficient

As shown in Fig. 3, when the DMR is at a high level (relative to the desired level), changes in the sampling period will affect the DMR significantly. The DMR could then be brought back to the desired level by only a small change in the sampling period. Accordingly, the value of $K$ should be set relatively small. Otherwise, when the DMR is at a low level, the effect of the change in sampling period on the DMR is less significant. To achieve the desired level of DMR more quickly, the value of $K$ should be set larger.

In this work, a simple yet illustrative algorithm given by (4) is used to adapt $K$.

$$K = \begin{cases} K_0/2 & \rho(k) > \rho_r + \Delta\rho^+ \\ K_0 & \rho_r - \Delta\rho^- \leq \rho(k) \leq \rho_r + \Delta\rho^+ \\ 2K_0 & \rho(k) < \rho_r - \Delta\rho^- \end{cases} \qquad (4)$$

where $K_0$ can be obtained from simulation experiments, $\Delta\rho^+$ and $\Delta\rho^-$ are user-specified parameters. Generally $K_0$ inversely relates to the slope of the curve of DMR at the operation point, and consequently changes with the transmission rate $r$ and the DMR setpoint $\rho_r$.

Besides the above equation, there are other advanced algorithms that could potentially be more efficient in adjusting $K$, for example, the gain scheduling method from adaptive control theory. However, these complex algorithms also add burdens to online computations associated with the feedback scheduler, thus causing larger overheads.

3.3.2. Deadline Miss Ratio Setpoint

For a given control system, the effects of sampling period and DMR on QoC are deterministic, while the DMR is related to the sampling period. Therefore, for a given system setup, there exists an optimal operating point, say ($h_r$, $\rho_r$), at which the system will in principle achieve the best QoC. Ideally, the best feedback scheduling performance could be achieved by setting the desired level of DMR to this optimal point. In practice, the relationships between the QoC, the DMR, and the sampling period are complicated, and therefore cannot be explicitly described. Most often a DMR setpoint close to the optimal one could be chosen through simulation and/or experimental studies.

Suppose the point A($h_{r1}$, $\rho_{r1}$) in the schematic diagram Fig. 4 is the setpoint for $r = 5.5$ Mbps, which is (very close to) the optimal operating point. When the transmission rate changes, e.g., from 5.5 to 11 Mbps, the operating point of the system will become B($h_{r2}$, $\rho_{r1}$) if a fixed DMR setpoint is used. Clearly, the sampling period at point B decreases relative to A. Since trade-offs should be made between DMR and sampling period so as to achieve the best QoC, it is still possible to improve the QoC relative to the operating point B by properly increasing the sampling period, which reduces the DMR. Therefore, if A is the optimal operating point for $r = 5.5$ Mbps, then the optimal sampling period for $r = 11$ Mbps will be some value, say $h_{r3}$, that falls in the interval ($h_{r2}$, $h_{r1}$). This suggests that when the transmission rate increases, the QoC could be improved through decreasing the value of $\rho_r$, for example, using $\rho_{r2}$ as the new setpoint.

**Figure 4.** Schematic diagram for adapting deadline miss ratio setpoint.

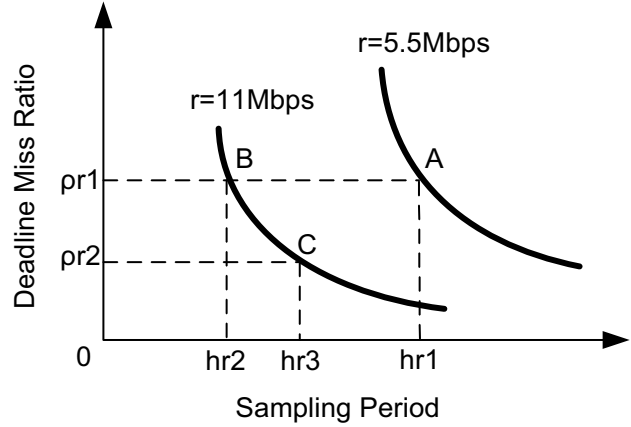

Based on this observation, we propose to adapt the DMR setpoint to different transmission rates. The setpoints used in this paper are simply set to:

$$\rho_r = \begin{cases} 10\% & \text{if } r=5.5 \\ 5\% & \text{if } r=11 \end{cases} \qquad (5)$$

Since only two cases, i.e. $r = 5.5$ and 11 Mbps, are considered in our simulation experiments, Equation (5) gives only the corresponding two values for $\rho_r$. Since practical control systems are always designed capable of tolerating some level of DMRs, there is often a considerably large room for choosing the value of $\rho_r$. Alternatively, compensation methods, e.g. [4], for packet losses can be adopted in control loops to alleviate the negative effect of deadline misses on QoC.

**Figure 5.** Pseudo code for cross-layer adaptive feedback scheduling.

```
Cross-Layer Adaptive Feedback Scheduling {
    //Determine h_max, K_0 at pre-runtime
    Measure deadline miss ratio ρ;
    Measure transmission rate r;
    //Adapt parameters if necessary
    IF r changes
        Update K_0
        Update ρ_r using (5)
    END
    Determine K using (4);
    //Compute new sampling periods
    Calculate e←ρ-ρ_r;
    Calculate Δh using (2);
    Reassign sampling periods according to (3);
}
```

Fig. 5 gives the pseudo code for the above-described feedback scheduling algorithm. It can be seen that this algorithm exhibits online adaptability in two aspects: 1) the adaptation of the proportional coefficient $K$ to deal with the nonlinear relationship between the DMR and the sampling period; 2) the adaptation of the DMR setpoint to deal with the changes in the transmission rate.

## 4. Event-Triggered Invocation

Feedback schedulers are usually time triggered. An obvious advantage of this mode is that it makes it convenient to design and analyse the feedback schedulers using feedback control theory and techniques. In wireless environments where the environmental changes including noise interference and node movement are irregular and bursty, however, it could be very difficult to choose an appropriate invocation interval for time-triggered feedback schedulers.

On one hand, the invocation interval cannot be set too small because accurate DMRs would be impossible to obtain. Therefore, the feedback scheduler with a large invocation interval will not be capable of coping with, in a timely fashion, interference and node movement that occur between two consecutive invocation instants. On the other hand, when a relatively small invocation interval is used, it is possible that the system stays in steady state for quite a long time, when there is actually no need for sampling period adjustment. In this situation, time-triggered feedback schedulers could potentially waste resources in periodic execution of feedback scheduling algorithms and unnecessary update of system parameters.

To address this problem, an event-triggered invocation mechanism is proposed to improve the efficiency of feedback schedulers. Discussed below is how to implement this mechanism.

*4.1. Design Methodology*

The schematic diagram of the event-triggered invocation mechanism is depicted in Fig. 6. With a structure similar to event-based controllers [31], there are basically two parts in this invocation mechanism [28], the event detector and the feedback scheduling algorithm. The event detector is time-triggered with a period of $T_{ED}$, while the feedback scheduling algorithm is triggered by the *execution-request* event issued by the event detector.

**Figure 6.** Schematic diagram of event-triggered invocation.

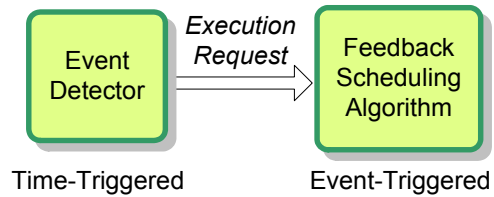

The design of the event detector is a key issue for implementing the event-triggered invocation mechanism. The major role of the event detector is to decide under what conditions the system needs to execute the feedback scheduling algorithm. Intuitively, when the DMR is in or close to a steady state, there is no need to execute the feedback scheduling algorithm. If the DMR deviates significantly from the desired level, it becomes mandatory to run the feedback scheduler to adjust system parameters. In this paper the following condition is used for issuing the *execution-request* event:

$$|\rho(k) - \rho_r| \geq \delta \qquad (6)$$

According to (6), the feedback scheduling algorithm will be executed if and only if the absolute difference between the actual DMR and its desired level is no less than a specific threshold $\delta$. In this way, the disadvantages of the time-triggered invocation mechanism with respect to response speed and overhead are avoided. Furthermore, the negative effect of measurement noise on the DMR may be alleviated naturally.

There are two important parameters, $T_{ED}$ and $\delta$. Generally speaking, choosing these parameters demands careful trade-offs between quick response and low overhead. Thanks to the small amount of computations of (6), it is possible to assign quite a small period $T_{ED}$ to the event detector to achieve quick response while keeping the feedback scheduling overhead small. The magnitude of the measurement noise should be taken into account when deciding the value of $\delta$. A $\delta$ value that is slightly bigger than the magnitude of measurement noise could be used to reduce the times of execution of the feedback scheduler, which results in smaller overheads.

## 5. Performance Evaluation

To evaluate the performance of the proposed event-triggered CLAFS scheme, this section conducts simulation studies for the case given in Section 2 using Matlab along with the TrueTime toolbox [12]. Consider the following two scenarios:
- Scenario I: The controller and the process are close to each other, WLAN operates at 11 Mbps, there is no interfering signals, $\delta = 0.03$, $K_0 = 0.018$;
- Scenario II: Due to increased distance between the controller and the process, the transmission rate drops to 5.5 Mbps, the interfering transmitter sends a data packet of 1 KB to the corresponding receiver every 10 ms, $\delta = 0.06$, $K_0 = 0.008$.

It can be seen that different $\delta$ values have been used in these two scenarios. This is because: 1) the DMR setpoints for different transmission rates are different, 2) this makes it convenient to compare the event-triggered feedback scheduling and time-triggered feedback scheduling, see Subsection 5.2.

Some parameters used in the simulations are as follows: the nominal sampling period $h_0 = 15$ ms, $h_{max} = 50$ ms, $T_{ED} = 500$ ms, $\Delta\rho^+ = 0.1$, and $\Delta\rho^- = 0.08$. It is worth mentioning that completely identical results cannot be guaranteed for each run of the simulation even with the same system setup. This is a natural consequence of the inherent stochastic feature of communications over WLAN. All results given below are the only representative ones among many obtained from a variety of simulation runs.

### 5.1. Feedback Scheduling vs. Traditional Design Method

In the first set of simulations, the proposed CLAFS method and the traditional design method without any feedback schedulers (denoted Non-FS) are compared. Since the three control loops are identical and WLAN adopts a random medium access control mechanism without distinguishing between them, all loops are equivalent in principle. Therefore, only the responses of one control loop are given below.

**Figure 7.** Control performance without feedback scheduling.

**Figure 8.** Control performance with CLAFS.

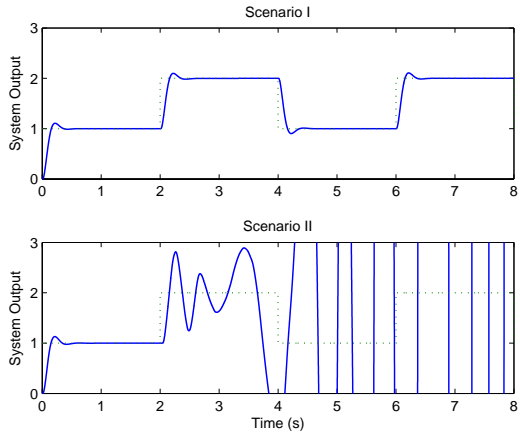
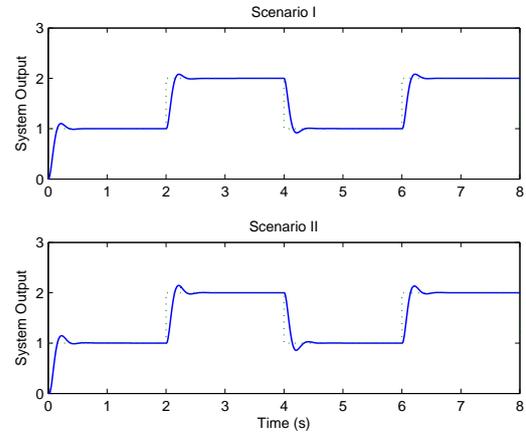

Fig. 7 depicts the step responses of loop 1 under different scenarios when the traditional method is used. It can be seen that the system performs quite well when the transmission rate of WLAN is 11 Mbps. However, under the second scenario, i.e., when the transmission rate drops to 5.5 Mbps with interfering signals, the system finally becomes unstable.

Fig. 8 shows the system performance when CLAFS is adopted. The system not only performs well under Scenario I, but also achieves good QoC under Scenario II.

The sampling periods and the DMRs under different schemes are shown in Figs. 9 and 10, respectively. With the traditional method, the sampling periods of all control loops are fixed at runtime, i.e., $h \equiv 15$ ms. When WLAN runs at 11Mbps (i.e., under Scenario I), the DMR is small with a mean of 0.9%. Consequently, the QoC is good. Under Scenario II, the DMR remains nearly 100% when time $t > 2s$, implying that almost all data packets transmitted on the WLAN miss their deadlines. This inevitably gives rise to system instability.

**Figure 9.** Sampling periods.

**Figure 10.** Deadline miss ratio.

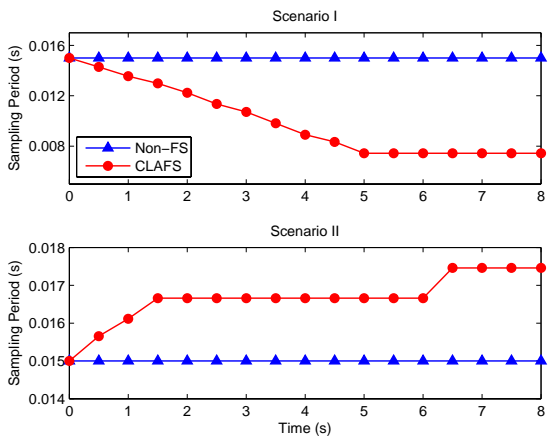
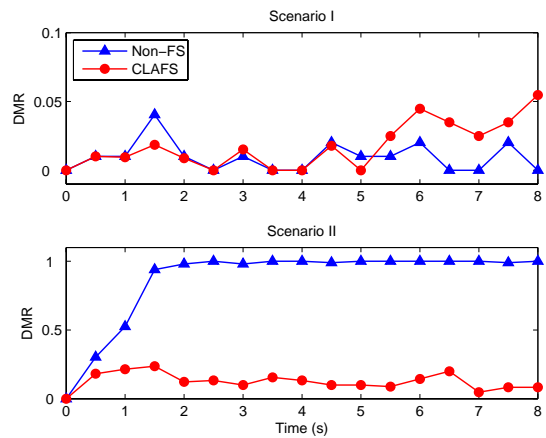

As can be seen from Figs. 9 and 10, the CLAFS scheme effectively controls the DMR through dynamically adjusting the sampling periods. Under Scenario I, the sampling periods of the control loops decrease from time t = 0, and remain at a steady level i.e. around 8 ms after time t = 5 s. The DMR also keeps at a low level, with a mean of 1.7%. Finally, it approaches its setpoint 5%. The levels of the DMR under both schemes are close, but the sampling periods are smaller when CLAFS is used.

Under Scenario II, CLAFS successfully avoids high DMRs by increasing the sampling periods gradually. After a transient process, the DMR keeps around the setpoint 10%. It can be seen that both the sampling periods and the DMR increase on average under Scenario II relative to Scenario I, which may have some negative effects on the QoC. Consequently, the QoC is slightly worse in Scenario II than in Scenario I, as shown in Fig. 8.

The above simulation results show that the proposed CLAFS scheme is able to effectively attack the problem of transmission rate changes and ambient noise interference, thus improving the quality of control of the whole system.

*5.2. Event-Triggered vs. Time-Triggered*

In the second set of simulations, the performance of event-triggered and time-triggered CLAFS methods is compared. To facilitate comparisons with the event-triggered scheme simulated in the first set of experiments, the invocation interval for the time-triggered feedback scheduler is set as $T_{FS} = T_{ED}$ = 500 ms.

Fig. 11 depicts the step responses of loop 1 under both scenarios when the time-triggered CLAFS scheme is applied. The QoC is pretty good. Comparing Fig. 11 with Fig. 8, it can be seen that the time-triggered and event-triggered CLAFS achieve comparable QoC.

**Figure 11.** Control performance with time-triggered feedback scheduling.

**Figure 12.** Sampling periods and deadline miss ratio for time-triggered feedback scheduling.

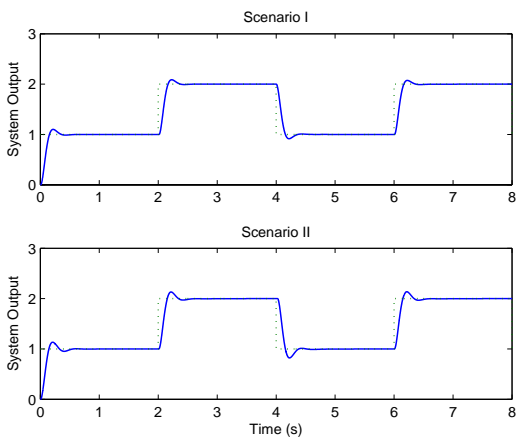
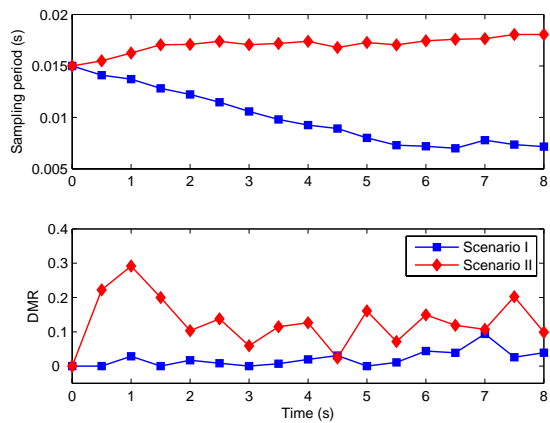

The sampling periods and the DMRs for the system using time-triggered CLAFS are presented in Fig. 12. They vary in the same manner as under event-triggered CLAFS. The main difference between them is that with event-triggered feedback scheduling the sampling periods remain unchanged at some consecutive sampling instants (see Fig. 9), which implies that the feedback scheduler does not actually execute, whereas the sampling periods are updated at every sampling instant when time-triggered feedback scheduling is used (see the upper part of Fig. 12).

To this point only the QoC is examined, which is *ideal* in that the effect of feedback scheduling execution is not taken into account. That is, in the above simulation experiments the overhead of feedback scheduling is neglected. For the purpose of comparison, the *times of execution* of the feedback scheduler is used as a simple criterion for measuring the feedback scheduling overhead.

**Table 1.** Comparison of event-triggered and time-triggered invocations.

|  | Scenario I | | Scenario II | |
| --- | --- | --- | --- | --- |
|  | Time-Triggered | Event-Triggered | Time-Triggered | Event-Triggered |
| ΣIAE | 1.131 | 1.127 | 1.295 | 1.293 |
| Times of Execution | 16 | 10 | 16 | 4 |

Table 1 compares the total control costs of the system (calculated by the sum of the integral of absolute error of each control loop) and the times of execution of the feedback scheduler with different invocation mechanisms. For different invocation mechanisms the overall QoC remains almost identical in both scenarios. In Scenario I, the times of execution of the feedback scheduler decreases 37.5% with event-triggered CLAFS as compared to the time-triggered case. In Scenario II it reduces from 16 to 4, with a relative reduction of 75.0%.

The above results show that the proposed event-triggered invocation mechanism yields significant reduction in feedback scheduling overheads while providing comparable feedback scheduling performance, thus improving the efficiency of the CLAFS scheme. Furthermore, by simply selecting a smaller $T_{ED}$ value, the event-triggered invocation mechanism can be used to achieve quicker response associated with the feedback scheduler, without introducing too large overheads.

## 6. Concluding Remarks

This paper deals with dynamic management of the communication resources in WCSs. A cross-layer adaptive feedback scheduling scheme, which features co-design of real-time control and wireless communications, has been developed. With this scheme, the effects of noise interference and changes in link capacity on QoC can be addressed effectively, thus enabling wireless control in dynamic and uncertain environments. To avoid the difficulty of time-triggered invocation in making trade-offs between response speed and overhead in wireless environments, an event-triggered invocation mechanism has also been proposed, which improves the practical performance of feedback scheduling.

The proposed scheme could be extended in several aspects. One possibility is generalizing the cross-layer design framework. For example, in order to take into account the effect of different MAC protocols, the MAC sub-layer may be included in the framework. In cases where the energy consumption of the nodes is a concern, physical-layer parameters such as the transmit power may be made available for other upper layers. Another possibility is improving the adaptive feedback scheduling algorithm. Given that the behaviour of the wireless network with respect to deadline miss ratio could be modelled with sufficient accuracy, for instance, it is possible to obtain analytically an optimized adaptive feedback scheduling algorithm using relevant control theory and techniques.

Our future work in this direction includes development of an experiment system for WCSs over WLAN, which will be used to assess the performance of the proposed scheme with more extensive results. Another topic is to conduct theoretical stability analysis of WCSs that employ the proposed feedback scheduling scheme.

## Acknowledgements

The first author would like to thank Prof Yu-Chu Tian at QUT, Australia, for valuable suggestions. This work is supported in part by China Postdoctoral Science Foundation under Grant No. 20070420232, Natural Science Foundation of China under Grant No. 60474064 and 60704024, Zhejiang Provincial Natural Science Foundation of China under Grant No. Y107476, Natural Science Foundation of Jiangsu under Grant No. BK2006573, and Australian Research Council (ARC) under Discovery Projects Grant No. DP0559111.